\documentstyle[psfig]{l-aa}

\def\LSUN{\hbox{$L_\odot$}}
\def\Lsun{\hbox{$L_\odot$}}
\def\HA{\hbox{H$\alpha$}}
\def\HB{\hbox{H$\beta$}}

\def\EW{\hbox{W$_\lambda$}}
\def\AV{\hbox{A$_{\rm V}$}}

\begin{document}

\title{
Spectropolarimetry of the Circinus galaxy
\thanks{Based on observations collected
at the European Southern Observatory, La Silla, Chile}}
\author{  E. Oliva\inst{1}, A. Marconi\inst{1,2,3}, A. Cimatti\inst{1}, 
S. di Serego Alighieri\inst{1} }
\institute{
            Osservatorio Astrofisico di Arcetri,
               Largo E.Fermi 5,
               I--50125 Firenze,
               Italy
	\and
            Dipartimento di Astronomia e Scienza dello Spazio,
            Largo E. Fermi 5,
            I--50125 Firenze, Italy
        \and
Present address: Space telescope Science Institute, 3700 S. Martin Drive,
Baltimore 21218 MD, USA
           }
\offprints{ E. Oliva}
\date{ Received 5 Sep; accepted 23 Sep 1997 }
\thesaurus{ ( 11.19.1; 11.09.1 Circinus ) }
\maketitle

\markboth{Oliva et al.: Spectropolarimetry of the Circinus galaxy}{}

\begin{abstract}
High quality 4500--6800 \AA\  
spectropolarimetric observations of the Circinus galaxy are reported.
These show polarized and
relatively broad (FWHM$\sim$3300 km/s) \HA\ (as well as marginal \HB)
arising from a $<$3\arcsec\ ($<$60 pc) region centered on the nucleus 
and coincident with the the peak of the narrow line region.
Broad \HA\ might also be present 
$\sim$8\arcsec\ (160 pc) SE from the nucleus and close to the dust
lane, but the result is only marginal and should be verified by 
higher s/n data.
The continuum is dominated by stellar emission
with $\le$3\% contribution from the scattered AGN continuum,
and {\it all the light} (i.e. all the stellar continuum and narrow lines) 
is polarized, most
probably by transmission through and scattering by dust in the disk.
The data are compatible with a simple model where the scattered
broad lines and AGN emission are quite polarized
(P$\sim$25\%) but account for only 1\% of the observed continuum.
The scattered broad \HA\  has an equivalent width \EW$\simeq$400 \AA,
a flux similar to the narrow \HA\  and
a luminosity $\simeq$5 $10^5$ \LSUN.
Assuming an efficiency of 1\% for the BLR mirror, the intrinsic
luminosity of the broad \HA\ is roughly  $5\,10^7$ \LSUN\  and
$\sim$0.5\% of the
FIR luminosity, a ratio similar to those found in
type 1 Seyferts.
Specific prediction for future observations
are also presented.

\end{abstract}
\section { Introduction }
After the discovery of polarized broad lines in NGC1068,
spectropolarimetry 
has
become a standard tool to trace  
hidden broad line regions (BLR)
in objects like type 2 Seyferts where   
BLR and AGN continuum are visually obscured 
(cf. Antonucci 1993 for a review).
In the standard model $\approx$1\% of the BLR photons
are scattered toward us by a '$\!$mirror' (dust or free electrons)
lying outside the obscuring 'torus', and the light appears partly
polarized.
Seyferts 2 with polarized BLR are usually characterized by
prominent narrow lines onto a mixture of stellar and
non stellar (featureless) continua, and the latter
accounts for a significant fraction ($>$10\%) of
the total observed continuum flux.
The narrow lines are much less polarized than the 
continuum and [OIII] is often (but not always) unpolarized.
The stellar flux is also basically 
unpolarized and is 
subtracted using suitable stellar templates to obtain the 'true'
polarization of the non stellar nuclear emission (e.g. Miller
\& Goodrich 1990).

The Circinus galaxy is an anonymous, nearby ($\simeq$4 Mpc) spiral
lying very close to the galactic disk ($b$=$-$3.8$^\circ$) in a region
of relatively low Galactic extinction (\AV$\simeq$1.5 magnitudes, 
Freeman et al. 1977).
Its Seyfert 2 activity was already suspected on the basis of IR colours
(Moorwood and Glass, 1984) and definitely demonstrated by recent
observations of optical and IR coronal lines (Oliva et al. 1994, hereafter O94,
Moorwood et al. 1996),
optical line images showing a spectacular [OIII] cone 
and \HA\ starburst ring (Marconi et al. 1994, hereafter M94), and
X--rays spectroscopy revealing a very prominent Fe-K line (Matt et al.
1996). Being very bright and $\sim$5 times closer than NGC1068, the 
Circinus galaxy
is the ideal laboratory to study the Seyfert 2 phenomenon.
The only drawback is that the object cannot be easily
studied in the UV because the central $\approx$3\arcsec\ regions suffer by 
significant local extinction (\AV$\ga$3 mag),
and this absorption strongly varies with position.
To our knowledge, no spectropolarimetric observations of this galaxy
exist. 

\begin{figure*}
\centerline{ \psfig{figure=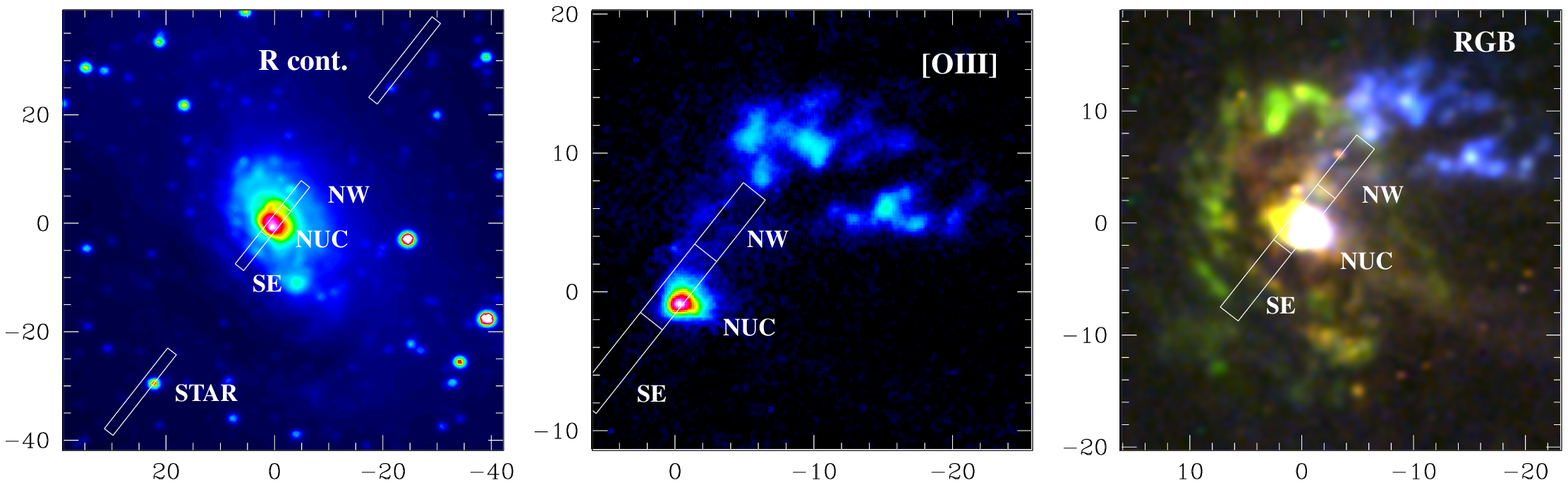,width=\hsize,angle=-90}}
  {\bf Fig. 1.}
  The position of the 2\arcsec\ slit 
  is overlaid onto a R frame taken with EFOSC1 (left panel), 
  a [OIII] line image (central panel) and on a 'true colour'
  (red=[SII], green=\HA+[NII], blue=[OIII]) representation of the Circinus
  galaxy. North is up, east is left and the numbers are arcsec 
  offsets from the nucleus.
  The line images are from M94 and are also available at
  http://www.arcetri.astro.it/$\sim$oliva.
\vskip10pt
\end{figure*}

This letter presents for the first time spectropolarimetric observations
of this galaxy. The observations and results are described in Sect.
2, 3 while Sect. 4 presents a simple model
which is also used to estimate the intrinsic properties of the
obscured BLR.

\section{The data}

The spectra were collected on March 7 1997 using EFOSC1 mounted
on the ESO 3.6m telescope. The detector was a Tek 512x512 CCD array
with 27$\mu$m (0.61\arcsec\ sky projected angle) pixels and the
instrument setup included
a rotating $\lambda$/2 plate and a Wollaston
prism in the collimated beam, and the disperser
%
%
was the low resolution ($\simeq$6.5 \AA/pixel) ESO B300 grism
%
%
%
The 2\arcsec\ broad slit was
aligned at PA=318$^\circ$, roughly along the [OIII] cone axis and
perpendicular to the galaxy disk (Fig.~1).
Twenty exposures with a total integration time of 5 hours were collected,
and consisted of
five cycles of four 15 minutes
exposures with the $\lambda/2$ plate rotated by 0, 22.5, 45 and 67.5
degrees.
Measurements of polarized (HD126593, HD298383),
unpolarized (HD64200) and spectroscopic (Hiltner 600) standard stars
were also performed for calibration purposes.
Standard reduction of the 2D frames was applied 
%
%
and three spectra were extracted at different positions along the slit (cf.
Fig.~1 and the caption of Fig.~2),
the total flux spectra are shown in Fig.~2. 
%
%

The polarimetric reduction of the 1D spectra was 
performed using the software written by J.R. Walsh under the MIDAS environment
and the resulting linear polarization degree ($P_{obs}$) and 
position angle ($\theta_{obs}$) are
displayed in Fig.~2.
These quantities must be corrected for the polarization
by our Galaxy whose polarization vector can be first estimated from
the spectra of the foreground star (Fig.~1) which yield
$P_{\rm V}$=1.8\% at $\theta$=68$^o$,
an angle equal to that found in the NW spectrum.
The corrected spectra are $P_{corr}$ and $\theta_{corr}$
which are also plotted in Fig.~2 where
the polarization degree of NW decreased to
$\simeq$0.8\% while its angle remained $\simeq$68$^o$
and equal to the galactic polarization angle. This indicates
that the NW spectrum may be intrinsically unpolarized 
and the correct amount of Galactic polarization could therefore be
%
%
$\simeq$2.6\% and larger than that suffered by the
foreground star which is probably too close to properly sample
the whole disk of our Galaxy.
Using $P_{\rm V}$=2.6\% the corrected spectra become $P'_{corr}$ and 
%
%
%
$\theta'_{corr}$
which are plotted in the bottom rows of Fig.~2.
The true polarization degree 
of the Circinus spectra is somewhere between
$P_{corr}$ and $P'_{corr}$.
Noticeably, the $P$ spectra
of the nucleus and of the SE region are rather
independent on the details of the correction for local 
polarization, i.e. $P_{corr}$ and $P'_{corr}$ are virtually equal in the
nuclear and SE spectra.
However, the polarization angle does depend on the correction applied
but is basically independent on wavelength in both $\theta_{corr}$
and $\theta'_{corr}$ (cf. Fig.~2).

\section{Results}

\subsection{The polarized broad \HA}

The most striking result is the prominent emission feature at
6575 \AA\ (observed $\lambda$) in the $P$ nuclear spectrum. The center
of this feature is within 100 km/s (1/10 of the instrumental resolution) 
of the wavelength of the narrow \HA\ (cf. O94), has a full 
width half maximum of 72 \AA\ (3300 km/s)
and is much broader than the unresolved \HA+[NII] complex in
the $F_\lambda$ spectrum which has FWHM=42 \AA.
The $P$ nuclear spectrum does not show other emission/absorption features with
the possible exception of \HB\ which is marginally (2$\sigma$)
detected at $\sim$4870~\AA.
In particular, no significant variation of $P$ and $\theta$
is visible at the position of the prominent [OIII] and [SII] narrow lines,
and this indicates that [NII] does not
contribute to the broad \HA, but rather dilutes it
(see also Sect. 4.2)

\begin{figure}
\centerline{\psfig{figure=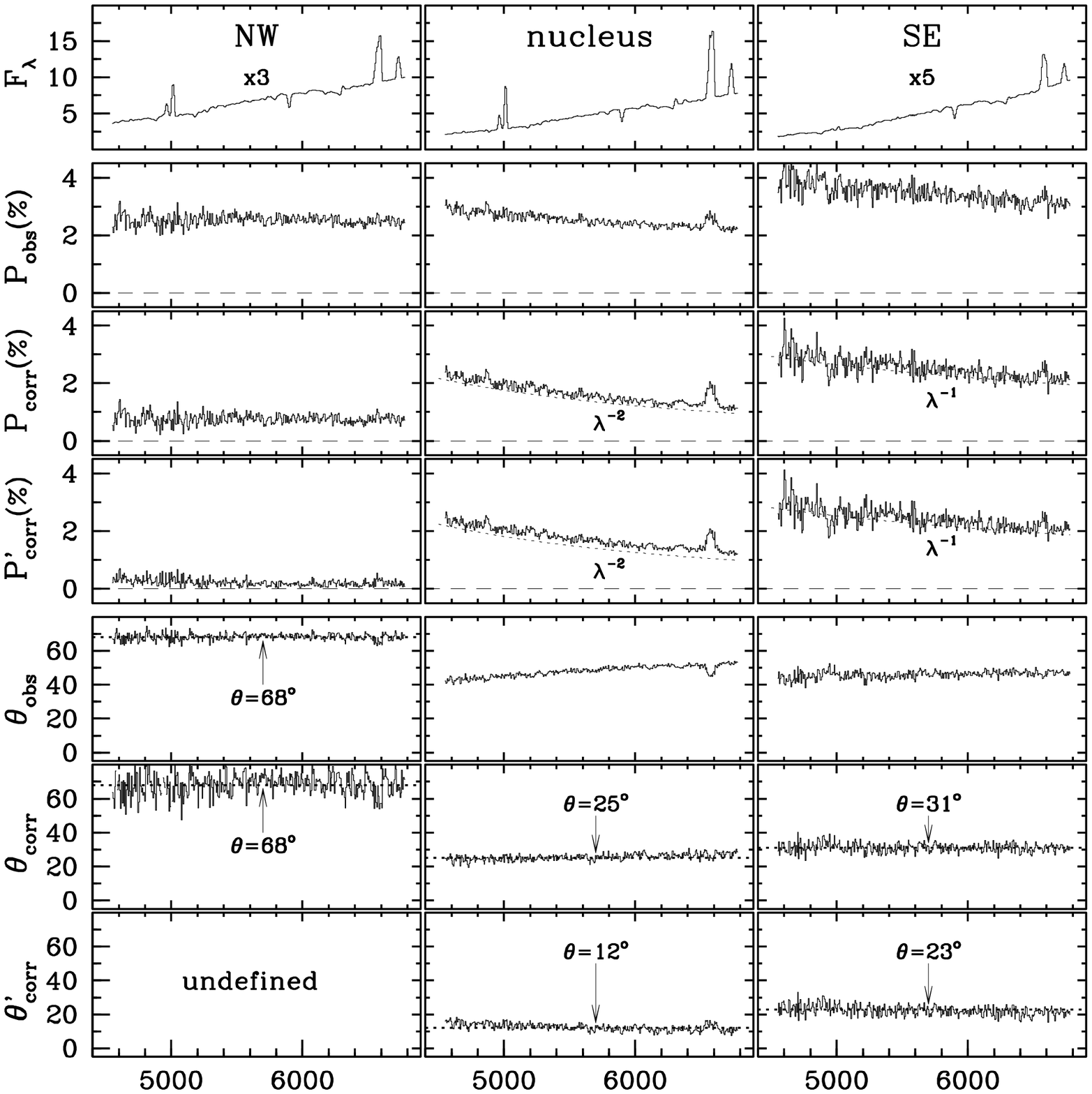,width=\hsize}}
{\bf Fig. 2.}
Extracted spectra along the slit (cf. Fig. 1),
'nucleus' is 2\arcsec x5\arcsec\ centered on the optical peak which is coincident
within 0.3\arcsec\ with the IR nucleus (M94), 'SE'
is 2\arcsec x6\arcsec\ centered 7\arcsec\ SE of the nucleus and toward the dust lane
(cf. Figs. 1, 2 of M94) and 'NW' is 2\arcsec x5\arcsec\ centered 5\arcsec\ NW of
the nucleus and within the [OIII] cone (cf. Fig. 1).
Wavelengths are in \AA\  and 
$F_\lambda$ is in units of 10$^{-15}$ erg cm$^{-2}$ 
s$^{-1}$ \AA$^{-1}$.
The observed linear polarization degree and angle are
 $P_{obs}$ and $\theta_{obs}$ while
$P_{corr}$, $\theta_{corr}$ and $P'_{corr}$, $\theta'_{corr}$
are the values corrected for local polarization using two different
estimates of the galactic polarization vector
(cf. Sect. 2).
Note that the spectra have the original resolution.
\end{figure}
The spatial extent of the broad \HA\  was estimated from a
row-by-row polarimetric reduction of the 2D spectra. 
The resulting distribution along the slit is peaked on the nucleus,
concentrated within $<$3\arcsec\  and basically unresolved.
Interestingly however, some broad \HA\ emission seems to reappear
$\sim$8\arcsec\ SE
of the nucleus, and this causes the marginal detection of broad \HA\ in the SE
$P$ spectrum (cf. Fig.~2). This indicates that 
another BLR mirror may exist $\ga$150 pc SE from the
nucleus and in the direction of the dust lane, but the evidence is only
marginal.  
Higher  s/n spectra are required to verify this possibility.

\subsection{Dilution of the scattered nuclear spectrum}
%
%
%

The standard method to determine the intrinsic polarization properties
of the nuclear spectrum is to estimate the contribution of stellar
flux to the observed continuum, and correct for it assuming that the 
diluting stellar continuum is unpolarized.
The stellar template normally used for this purpose
is that of an old stellar system (e.g. M32), but this provides a
very poor fit
to the Circinus galaxy whose spectrum shows strong \HB\
absorption and other features typical of 
B-A stars associated to a relatively young (circum)nuclear starburst
whose presence is also demonstrated by other observational evidences
(e.g. Oliva et al. 1995).

\begin{figure}
\centerline{\psfig{figure=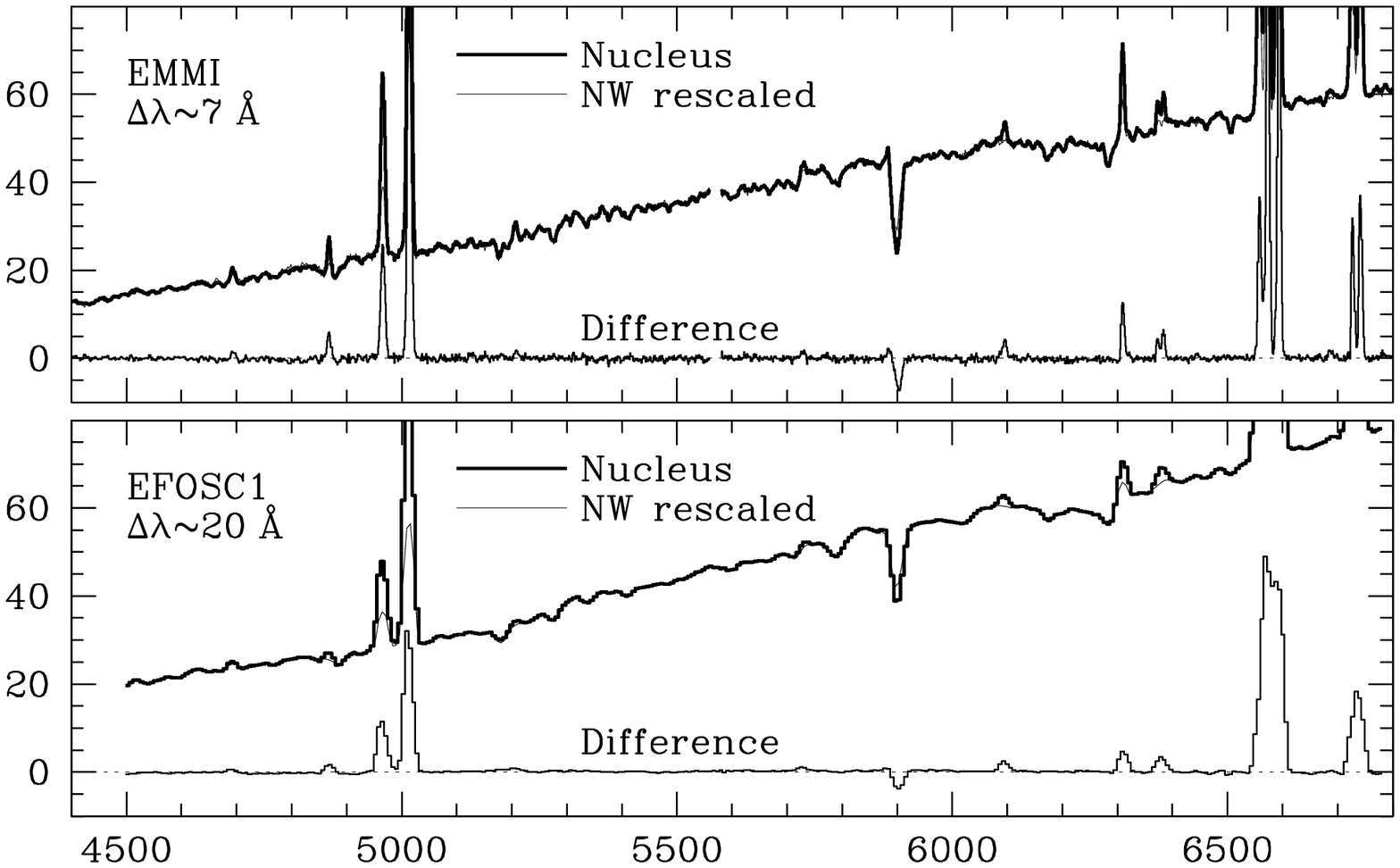,width=\hsize}}
{\bf Fig. 3.}
Comparison between the observed nuclear and NW spectra, the latter is
scaled and reddened to match the continuum level and shape of the
nuclear spectrum. The lower panel shows the EFOSC1 spectra discussed
here while the upper panel is based on higher resolution EMMI observations
which will be discussed elsewhere. 
Note that the narrow emission lines
and interstellar Na-D absorption are relatively stronger in the nucleus, but
the stellar absorption features have the same equivalent widths
in both spectra which are virtually indistinguishable.
This indicates that the scattered featureless
nuclear continuum accounts for a very small fraction ($<$3\%)
of the observed continuum.
\end{figure}

A much more accurate, and indeed the most natural stellar template 
is the NW spectrum which is
intrinsically unpolarized
and does not show any trace of the scattered nuclear component.
A comparison between the nuclear and the NW spectra is shown in Fig.~3
where the most remarkable result is that the equivalent width of the
stellar features is the same in both spectra, and the maximum contribution
from a featureless continuum is only $<$3\% of the total
observed flux.
A straight correction for the stellar dilution is therefore impossible.

\section{Discussion}
\begin{figure*}
\centerline{
  \psfig{figure=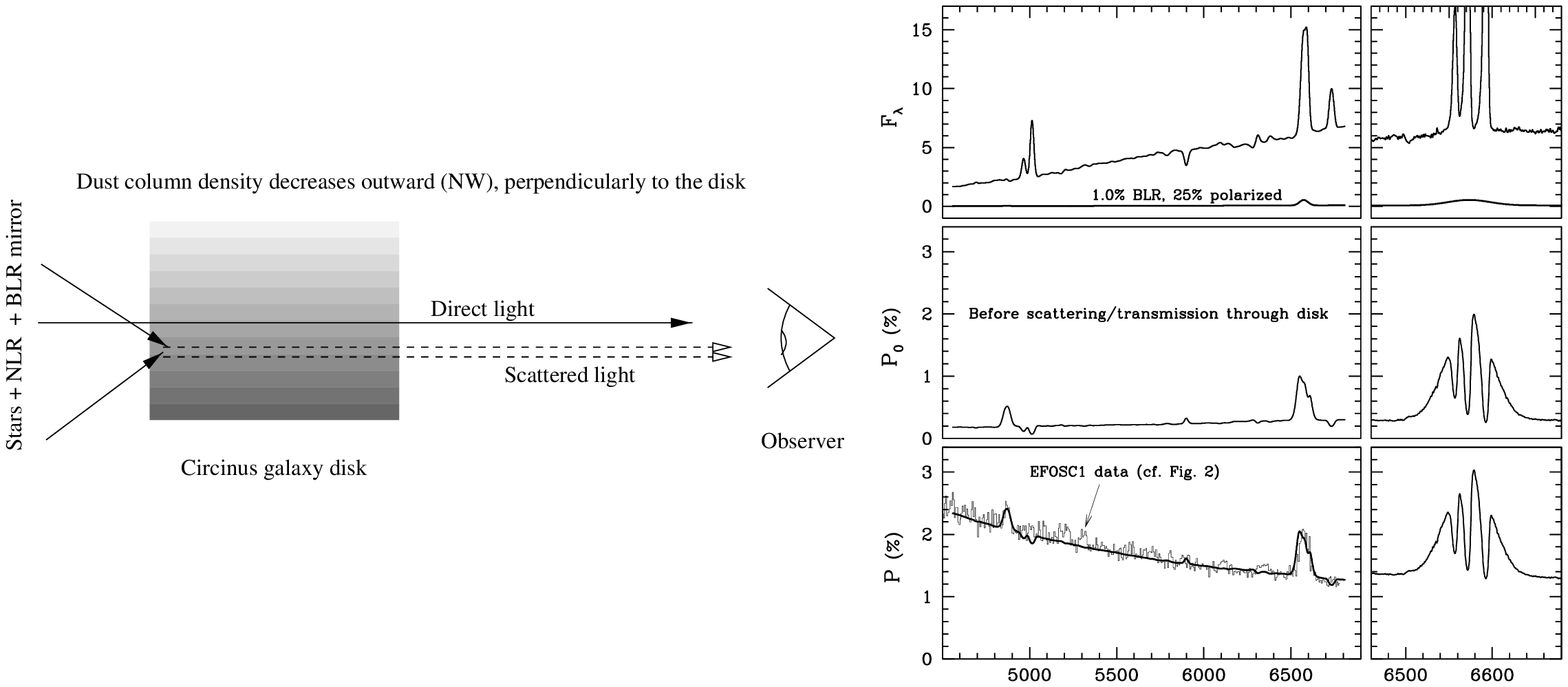,width=13.7cm,angle=-90}\hfill
\parbox[b]{4.0cm}
{
  {\bf Fig. 4.}
  Left: sketch of the model used to interpret the observed
  polarization of the stellar and NLR spectra. The observer mostly
  sees direct (extincted) radiation plus a small fraction of off-axis 
  light scattered toward the line of sight by dust in the disk.
  Right: results of the model discussed in Sect. 4.2, the
  complete spectra are at the EFOSC1 resolution while the
  details around \HA\ are the predicted spectra at resolving power $R$=2000.
\vskip5pt
}}
\end{figure*}
\subsection{Origin of the continuum polarization}

The observed  $\sim$2\% continuum polarization  
is very difficult to interpret in terms of
scattered nuclear light
diluted by unpolarized stellar emission because this would require
an unreasonably large ($>$70\%) intrinsic polarization.
Besides, in such a scenario the $P$ spectra should
contain narrow [OIII], [SII] 
absorption, unless
the NLR intrinsic polarization is finely tuned to avoid this effect.
%
%
 
The most  natural interpretation 
is that {\it all the light} (i.e. stars, NLR and scattered
BLR) is weakly polarized by either transmission through, or single scattering
by the dust in the galactic disk. A highly simplified
though plausible  scenario is that sketched
in Fig.~4 where the observer sees the extincted radiation from the stars, the
NLR and the BLR mirror, plus a small fraction of 'off-axis' light
scattered toward the line of sight by the dust in the disk of the
Circinus galaxy. A non-zero polarization is ensured by the
asymmetry of the dust distribution whose
column density increases toward SE producing the observed sharp extinction
gradient, and by the non uniform distribution of the stellar and NLR
emission which is strongly peaked on the nucleus (cf. M94).
The shape of the observed $P$ spectrum depends on the relative contribution
of scattering and transmission polarizations, and the latter is more
important in the highly extincted SE region thus producing a flatter $P$
spectrum.

The fact that the corrected polarization angle is basically constant with 
$\lambda$ and does not show significant variations within the broad \HA\
profile requires 
that the polarization by the
BLR mirror and scattering/transmitting dust have similar
position angles. This is not unreasonable because the corrected $\theta$ 
is $\sim$20$^\circ$, roughly parallel to the disk and dust lane 
and perpendicular to the axis of the [OIII] cone
(cf. Figs. 1, 2 of M94). If the BLR mirror is along the cone axis and the
magnetic field is aligned with the galaxy disk, it naturally follows that
all polarization angles should be similar.

\subsection{The physical properties of the BLR and its mirror }

A detailed treatment of the radiation transfer and scattering through
the disk 
is beyond the aims of this paper. 
Here we concentrate on the properties
of the BLR scattered spectrum and make the reasonable
assumption that the polarization introduced by transmission/scattering
through the disk could be parameterized by 
$P_{disk}\approx (\lambda/5500)^{-\alpha}$
where the slope $\alpha$ is $\ga$2 when single scattering by small
grains dominates, while is $\simeq$0 when the main polarization
mechanism is transmission through aligned non-spherical grains.
If the polarization angles are similar, and as long as the polarization
degrees are small (which is our case), the observed degree of linear
polarization is simply the sum of $P_{disk}$ and $P_0$, the polarization 
observed by ideally removing the galaxy disk.
We model $P_0(\lambda)$ assuming that the nuclear scattered spectrum is
polarized, while both the stellar continuum and narrow lines are unpolarized.

The results of a toy model with a 1\% scattered AGN
are shown in right hand panel of Fig.~4.
Due to the large dilution, the value of $P_0$ is very small at all wavelengths
but at the positions of the broad \HA\ whose
amplitude is a factor $\sim$5 the nuclear continuum level and 
therefore stands out
in $P_0$ because it is $\sim$5 times less diluted than the surrounding
continuum, and the same applies to the weaker \HB.
The narrow lines appear in absorption because they further dilute the
scattered spectra, but their amplitude is very small simply because
the the continuum $P_0$ is very low.
The effect of the disk polarization is simply to add a smooth continuum
to $P_0$, and does not affect the amplitude of the emission (BLR)
and absorption (NLR) lines.
Details of the model of Fig.~4 are as follows.\\
-- Nuclear continuum:  
   $I_\lambda\propto\lambda^{-1}$ scattered by a gray mirror\\
-- \EW(\HA-broad)=400 \AA \ \ \ ; \ \ \ I(\HA-broad)/I(\HB-broad) = 3.5 \\
-- BLR mirror extincted by \AV=5 mag (same as NLR, 
cf. O94) \\
-- Scattered spectrum has $P$=25\% at all wavelengths\\
-- Nuclear scattered continuum is 1\% of observed $F_\lambda$ \\
-- Stellar and NLR spectra are from high resolution EMMI data \\
-- Polarization by the disk: 
   $P_{disk}(\lambda)=0.0148$\,$(\lambda/5500)^{-2}$\\
These values are by no
means unique because similarly good fits could be obtained by e.g.
decreasing the intrinsic equivalent width of \HA\  and increasing the
contribution by the BLR to the total observed continuum, 
but the fit rapidly deteriorates for equivalent widths below 200 \AA\
and \EW(\HA)$>$150 \AA\  could be considered a tight lower limit.

A quite firm result is that the scattered light cannot
be much bluer than assumed otherwise the broad \HB\ would appear
too strong.
This sets a tight lower limit \AV$>$2 for the extinction suffered 
by the BLR mirror,
but gives no useful information on the
intrinsic spectral efficiency of the mirror whose extinction is unknown.
In other words, the scattered light could equally well come from
a 'gray mirror' suffering an extinction similar to the NLR (the
assumption of Fig.~4) or from a 'blue mirror' suffering a larger
exctinction.  

The best constrained parameter is the observed flux of broad \HA\
which is $3\, 10^{-14}$ erg cm$^{-2}$ s$^{-1}$ and translates into
$5\, 10^5$ \Lsun\  if the extinction toward the BLR mirror is similar
to that measured for the narrow lines. Assuming a 'standard' mirror
efficiency of 1\%, the intrinsic luminosity of broad \HA\  becomes
$5\, 10^7$ \Lsun\  or 0.5\%  of the IRAS FIR luminosity,
a ratio similar to those found in many type 1 Seyferts (see e.g. 
Table 1 of Ward et al. 1988).

Interesting predictions of the model are the detection of
circular polarization due to scattering of the linearly polarized BLR,
and high resolution $P$ spectra which
should reveal a stronger \HA\ broad  component
with sharp absorption features at the positions
of the narrow [NII] and \HA\ lines (Fig.~4)

%
%


%

\end{document}